\date{281206}
\documentclass[showpacs,twocolumn]{revtex4}
\usepackage{graphicx,psfrag,amsmath,amssymb,amsfonts,bbm,latexsym,color,dcolumn,
epsf,graphpap}

\definecolor{red}{rgb}{1,0,0}
\definecolor{blue}{rgb}{0,0,1}
\definecolor{skyblue}{rgb}{0,0,.5}
\definecolor{green}{rgb}{0,1,0}
\definecolor{orange}{cmyk}{0,.4,1,0}

\begin{document}
\title{Decoherence of domains and defects at phase transitions}

\author{F.C. Lombardo$^a$ \footnote{lombardo@df.uba.ar}}
\author{R.J. Rivers$^b$ \footnote{r.rivers@imperial.ac.uk}}
\author{P. I. Villar$^a$ \footnote{paula@df.uba.ar}}

\affiliation{$^a$Departamento de F\'\i sica {\it Juan Jos\'e
Giambiagi}, FCEyN UBA, Facultad de Ciencias Exactas y Naturales,
Ciudad Universitaria, Pabell\' on I, 1428 Buenos Aires, Argentina}

\affiliation{$^b$Theoretical Physics Group; Blackett Laboratory,
Imperial College, London SW7 2BZ - UK}

\date{today}
\begin{abstract}
In this further letter on the onset of classical behaviour in
field theory due to a phase transition, we show that it can be
phrased easily in terms of the decoherence functional, without
having to use the master equation. To demonstrate this, we
consider the decohering effects due to the displacement of domain
boundaries, with implications for the displacement of defects, in
general. We see that decoherence arises so quickly in this event,
that it is negligible in comparison to decoherence due to field
fluctuations in the way defined in our previous papers.
\end{abstract}


\maketitle

The standard big bang cosmological model of the early universe,
with its period of rapid cooling, gives a strong likelihood of
phase transitions, with concomitant symmetry breaking. This paper
is a further paper in a sequence by ourselves and collaborators
\cite{npb,lomplb,proceed1} in which we explore the way in which
such phase transitions naturally take us from a quantum to
classical description of the universe.

That (continuous) transitions can lead rapidly to classical
behaviour is not surprising. Classical behaviour has two
attributes. i) {\it Classical correlations}: By this is meant that
the Wigner function(al) $W[\pi ,\phi]$ peaks on classical
phase-space trajectories. ii) {\it
 Diagonalisation} of the
 decoherence functional, whose role is to
describe consistent histories.

Continuous transitions supply both ingredients. Firstly, the field
ordering after such a transition is due to the growth in amplitude
of unstable long-wavelength modes, which arise automatically from
unstable maxima in the potential. From the papers of Guth and Pi
\cite{guthpi} onwards, it has been appreciated that unstable modes
lead to correlations through squeezing, and we shall not consider
it further. Secondly, we understand diagonalisation to be an
almost inevitable consequence of coarse-graining. The stable
short-wavelength modes of the field \cite{lombmazz}, together with
the other fields with which it interacts, form an environment
whose coarse-graining enforces diagonalisation and makes the
long-wavelength modes decohere. It is how this is implemented that
is the basis of this letter.

We stress that there is more than one way to formulate
diagonalisation. In our earlier papers \cite{npb,lomplb,proceed1}
we required the density matrix $\rho
 (t)$ itself to become diagonal, rather than the decoherence
 functional. Whichever approach we adopt, there is an issue as to which field basis
 we attempt to enforce diagonalisation, which can only be approximately achieved.
 For an infinite degree of freedom system field, ideally we should think of diagonalisation
 functionally. In practice this is impossible to achieve, and we are forced
 to adopt a piecemeal approach in which we
 make 'mini-superspace' approximations in which a finite number of degrees of freedom
 are isolated as the most significant. On comparing these, the relevant ones are those
 which decohere last. Since we are looking for the onset of classical
 behaviour, we take adjacent classical solutions with which to require no quantum interference.
  With these caveats we  shall see later that the two approaches lead to the same results.
In each case a probabilistic description
  is obtained, but the approach given here permits easier
  calculation.

  Since
  phase transitions take place in a finite time, causality
guarantees that correlation lengths remain finite. In
  our previous work, and this, we use the formation
  of domains after a transition to characterise the onset of classicality.
  To recapitulate, we showed \cite{npb,lomplb,proceed1},
  by using the master equation for the reduced density matrix,
  that the environment renders the
long-wavelength modes of the order parameter field classical at
early times, by or before the transition is complete. In
particular, those modes on the scale of the domain size will have
decohered, even though the modes on the scale of domain boundary
thickness do not.  In this paper we recreate that result
  rather more simply and show that a parallel result holds due to
  small displacements of domain boundaries, in a coarse-graining
  that is insensitive to their positions.

  This latter result has another consequence.  If the symmetry
breaking permits nontrivial homotopy groups the frustration of the
order parameter fields is resolved by the creation of topological
defects to mediate between the different ground states
\cite{kibble1,zurek1}. Since defects are, in principle,
observable, they provide an excellent experimental tool for
determining how phase transitions occur. For the simple  theory
that we shall consider here, that of a real scalar field with
double-well potential, the domain boundaries {\it are} the defects
(domain walls) and we can view our results as decoherence as
induced by small displacements of defects. The generalization to a
complex field $\phi$ is straightforward, and has been considered
elsewhere \cite{lomplb2}. This gives more substance to our
preliminary attempts  to show that vortex defects are also
classical by the time of their production \cite{lomplb2}.

As in our earlier work, we restrict ourselves to flat space-time.
The extension to non-trivial metrics is straightforward in
principle \cite{POFer}.

We now consider the case of a real quantum field $\phi$ with
double-well potential in detail. As we have said, the field
ordering after the transition begins is due to the growth of
long-wavelength modes. For these modes the environment consists of
the short-wavelength modes of the field, together with all the
other fields with which $\phi$ inevitably interacts in the absence
of selection rules \cite{huzhang,lombmazz}. The inclusion of
explicit environment fields is both a reflection of the fact that
a scalar field in isolation is physically unrealistic, as well as
providing us with a systematic approximation scheme \cite{npb}. To
be specific, the simplest classical action with scalar and
environmental fields $\chi_{\rm a}$ is
\begin{equation}
S[\phi , \chi ] = S_{\rm syst}[\phi ] + S_{\rm env}[\chi ] +
S_{\rm qu}[\phi ,\chi ], \label{action0}
\end{equation}
where (with $\mu^2$, $m^2 >0$ )
\begin{equation}
S_{\rm syst}[\phi ] = \int d^4x\left\{ {1\over{2}}\partial_{\mu}
\phi\partial^{\mu} \phi + {1\over{2}}\mu^2 \phi^2 -
{\lambda\over{4}}\phi^4\right\}, \nonumber \end{equation}
\begin{equation}
S_{\rm env}[\chi ] = \sum_{\rm a=1}^N\int d^4x\left\{
{1\over{2}}\partial_{\mu}\chi_{\rm a}
\partial^{\mu}
\chi_{\rm a} - {1\over{2}} m_{\rm a}^2 \chi^2_{\rm a}\right\},
\nonumber
\end{equation}
and the most relevant interactions between system and environment
are of the biquadratic form
\begin{equation}
S_{\rm qu}[\phi ,\chi ] = - \sum_{\rm a=1}^N\frac{g_{\rm a}}{8}
\int d^4x ~ \phi^{\rm 2} (x) \chi^{\rm 2}_{\rm a} (x).
\label{Sint}
\end{equation}
Even if there were no external $\chi$ fields with a quadratic
interaction of kind of Eq.(\ref{Sint}), the interaction between
short and long-wavelength modes of the $\phi$-field can be recast,
in part, in this form, showing that such a term is obligatory.

Although the system field $\phi$ can never avoid the decohering
environment of its own short-wavelength modes \cite{lombmazz}, to
demonstrate the effect of an environment we first consider the
case in which the environment is taken to be composed only of the
fields $\chi_{\rm a}$. Since environments have a cumulative effect
on the onset of classical behaviour, the inclusion of a further
component of the environment {\it reduces} the time it takes for
the system to behave classically. Thus it makes sense to include
the environment one part after another, since we can derive an
{\it upper} bound on that time at each step.

We have shown elsewhere \cite{npb,lomplb,proceed1} that, in order
to make our calculations as robust as possible, we need a
significant part of the environment to have a strong impact upon
the system-field, but not vice-versa. The simplest way to
implement this is to take a large number $N\gg 1$ of scalar
$\chi_{\rm a}$ fields with comparable masses $m_{\rm a}\simeq \mu$
weakly coupled to the $\phi$, with $\lambda$, $g_{\rm a} \ll 1$
(for details see Ref.\cite{npb}). Thus, at any step, there are $N$
weakly coupled environmental fields influencing the system field,
but only one weakly self-coupled system field to back-react upon
the explicit environment.

 For
one-loop consistency it is sufficient, at order of magnitude
level, to take identical $g_{\rm a} = g/\sqrt{N}$.  Further, at
the same order of magnitude level, we take $g\simeq\lambda$. This
is very different from the more usual large-N $O(N+1)$-invariant
theory with one $\phi$-field and $N$ $\chi_{\rm a}$ fields,
dominated by the $O(1/N)$ $(\chi^2)^2$ interactions, that has been
the standard way to proceed for a {\it closed} system \cite{boya}.
With our choice there are no direct $\chi^4$ interactions, and the
indirect ones, mediated by $\phi$ loops, are depressed by a factor
$g/\sqrt{N}$. In this way the effect of the external environment
qualitatively acts as a proxy for the effect of the internal
environment provided by the short-wavelength modes of the
$\phi$-field, but in a more calculable way.

We shall assume that the initial states of the system and
environment are both thermal, at a high temperature $T_{0}>T_{\rm
c}$. We then imagine a change in the global environment (e.g.
expansion in the early universe) that can be characterised by a
change in temperature from $T_{0}$ to $T_{\rm f}<T_{\rm c}$. That
is, we do not attribute the transition to the effects of the
environment-fields.

Given our thermal initial conditions it is not the case that the
full density matrix has $\phi$ and $\chi$ fields uncorrelated
initially, since it is the interactions between them that leads to
the restoration of symmetry at high temperatures. Rather, on
incorporating the hard thermal loop 'tadpole' diagrams of the
$\chi$ (and $\phi$) fields in the $\phi$ mass term leads to the
effective action for $\phi$ quasiparticles,
\begin{equation}
S^{\rm eff}_{\rm syst}[\phi ] = \int d^4x\left\{
{1\over{2}}\partial_{\mu} \phi\partial^{\mu} \phi - {1\over{2}}
m_{\phi}^2(T_0) \phi^2 - {\lambda\over{4}}\phi^4\right\}
\label{Stherm}
\end{equation}
where, in a mean-field approximation, $m_{\phi}^2(T_0)= -\mu^2
(1-T_0^2/T_{\rm c}^2)$ for $T\approx T_{\rm c}$. As a result, we
can take an initial factorised density matrix at temperature $T_0$
of the form ${\hat\rho}[T_0] = {\hat\rho}_{\phi}[T_0] \times
{\hat\rho}_{\chi}[T_0]$, where ${\hat\rho}_{\phi}[T_0]$ is
determined by the quadratic part of $S^{\rm eff}_{\rm syst}[\phi
]$ and ${\hat\rho}_{\chi}[T_0]$ by $S_{\rm env}[\chi_{\rm a} ]$.
That is, the many $\chi_{\rm a}$ fields have a large effect on
$\phi$, but the $\phi$-field has negligible effect on the
$\chi_{\rm a}$.

Provided the change in temperature is not too slow the exponential
instabilities of the $\phi$-field grow so fast that the field has
populated the degenerate vacua well before the temperature has
dropped significantly below $T_{\rm c}$ \cite{moro}. Since the
temperature $T_{\rm c}$ has no particular significance for the
environment fields, for these early times we can keep the
temperature of the environment fixed at $T_{\chi}\approx T_{\rm
c}$ (our calculations are only at the level of orders of
magnitude).  Meanwhile, for simplicity the $\chi_{\rm a}$ masses
are fixed at the common value $ m\simeq\mu$.

It is sufficient for our purposes here to take an instantaneous
quench.  Slower quenches make the analytic calculations very much
more difficult, without changing the qualitative nature of the
results \cite{npb}.

The notion of consistent histories provides an alternative
approach to classicality to trying to solve the master equation,
as we have done previously. Quantum evolution can be considered as
a coherent superposition of fine-grained histories. Since we need
to be able to distinguish different classical system-field
configurations evolving after the transition, we work in the
field-configuration basis. If one defines the c-number field $\phi
(x)$ as specifying a fine-grained history, the quantum amplitude
for that history is $\Psi [\phi] \sim e^{iS[\phi]}$ (we work in
units in which $\hbar =1$).

In the quantum open system approach that we have adopted here, we
are concerned with coarse-grained histories
\begin{equation}
\Psi [\alpha] = \int {\cal D}\phi ~ e^{iS[\phi]}\alpha [\phi]
\end{equation}
where $\alpha [\phi]$ is the filter function that defines the
coarse-graining. In the first instance this filtering corresponds
to tracing over the $\chi_a$ degrees of freedom.

>From this we define the decoherence functional for two
coarse-grained histories as
\begin{equation}
 {\cal D}[\alpha^+,\alpha^-] = \int {\cal D}\phi^+{\cal
 D}\phi^-~e^{i(S[\phi^+]-S[\phi^-])}\alpha^+ [\phi^+]\alpha^-
 [\phi^-].
\end{equation}
${\cal D}[\alpha^+,\alpha^-]$ does not factorise because the
histories $\phi^{\pm}$ are not independent; they must assume
identical values on a spacelike surface in the far future.
Decoherence means physically that the different coarse-graining
histories making up the full quantum evolution acquire individual
reality, and may therefore be assigned definite probabilities in
the classical sense.

A necessary and sufficient condition for the validity of the sum
rules of probability theory (i.e. no quantum interference terms)
is \cite{Gri}
\begin{equation}
 {\rm Re}{\cal D}~[\alpha^+,\alpha^-]\approx 0,
\end{equation}
when $\alpha^+\neq\alpha^-$ (although in most cases the stronger
condition ${\cal D}[\alpha^+,\alpha^-]\approx 0$ holds
\cite{Omn}). Such histories are consistent \cite{GH}.

For our particular application, we wish to consider as a single
coarse-grained history all those fine-grained ones where the full
field $\phi$ remains close to a prescribed classical field
configuration $\phi_{\rm cl}$. The filter function takes the form
\begin{equation}
 \alpha_{\rm cl}[\phi ] = \int {\cal D}J~
 e^{i\int J(\phi - \phi_{\rm cl})}\alpha_{\rm cl}[J].
\end{equation}
In the general case, $\alpha[\phi]$ is a smooth function (we
exclude the case $\alpha[\phi]=$ const, where there is no
coarse-graining at all). Using

\begin{equation}J\phi \equiv \int d^4x J(x) \phi (x),
\end{equation}
we may write the decoherence functional between two classical
histories as

\begin{eqnarray}
{\cal D}[\alpha^+,\alpha^-] &=& \int {\cal D}J^+{\cal
 D}J^-~e^{i W[J^+,J^-] - (J^+ \phi_{\rm cl}^+ - J^- \phi_{\rm cl}^-)}\nonumber \\
&\times & \alpha^+[J^+]\alpha^{-*}[J^-],
\end{eqnarray}
where

\begin{equation}
e^{i W[J^+,J^-]} = \int  {\cal D}\phi^+ {\cal D}\phi^- ~
e^{i(S[\phi^+] - S[\phi^-] + J^+\phi^+ - J^-\phi^-)}
,\end{equation} is the closed-path-time generating functional
\cite{calhu}.

In principle, we can examine adjacent general classical solutions
for their consistency but, in practice, it is simplest to restrict
ourselves to particular solutions $\phi^{\pm}_{\rm cl}$, according
to the nature of the decoherence that we are studying. Initially,
as we said earlier, we have made a de facto separation into the
order parameter field $\phi$ and its explicit environmental fields
$\chi_a$ whereby, in a saddle-point approximation over $J$,
\begin{equation}
 {\cal D}[\phi^+_{\rm cl},\phi^-_{\rm cl}] \sim
 F[\phi^+_{\rm cl},\phi^-_{\rm cl}].
\end{equation}
 $F[\phi^+, \phi^-]$ is the Feynman-Vernon \cite{feynver} influence functional (IF) (see
Ref. \cite{lombmazz,npb} for details). The influence functional is
written in terms of the influence action $A[\phi^+_{\rm
cl},\phi^-_{\rm cl}]$ as
\begin{equation}
  F[\phi^+_{\rm cl},\phi^-_{\rm cl}]=
 \exp \{iA[\phi^+_{\rm cl},\phi^-_{\rm cl}]\}.
\end{equation}
As a result,
 \begin{equation}
|{\cal D}[\phi^+_{\rm cl},\phi^-_{\rm cl}]| \sim
 \exp \{-{\rm Im}\delta A[\phi^+_{\rm cl},\phi^-_{\rm cl}]\},
 \end{equation}
where $\delta A$ is the contribution to the action due to the
environment.

>From this viewpoint, once we have chosen the classical solutions
of interest, adjacent histories become consistent at the time
$t_D$, for which

\begin{equation} 1\approx {\rm Im~\delta A}\vert_{t = t_D}. \label{tD2}
\end{equation}

As we are considering weak coupling with the environment fields,
we may expand the influence functional $F[\phi^+, \phi^-]$ up to
second non-trivial order in coupling strengths for large $N$.
Higher terms are depressed by powers of $N$. The general form of
the influence
action is then \cite{lombmazz,calhumaz} %
\begin{eqnarray}
\delta A[\phi^+,\phi^-] &= &\{\langle S_{\rm
int}[\phi^+,\chi^+_{\rm a}]\rangle_0 - \langle S_{\rm
int}[\phi^-,\chi^-_{\rm a}]\rangle_0\} \nonumber
\\ &&+{i\over{2}}\{\langle S_{\rm int}^2[\phi^+,\chi^+_{\rm a}]\rangle_0 - \big[\langle
S_{\rm int}[\phi^+,\chi^+_{\rm a}]\rangle_0\big]^2\}\nonumber
\\ &&- i\{\langle S_{\rm int}[\phi^+,\chi^+_{\rm a}] S_{\rm int}[\phi^-,\chi^- _{\rm a}]
\rangle_0 \nonumber \\&-& \langle S_{\rm int}[\phi^+,\chi^+_{\rm
a}]\rangle_0\langle S_{\rm
int}[\phi^-,\chi^-_{\rm a}]\rangle_0\} \label{inflac} \\
&&+{i\over{2}}\{\langle S^2_{\rm int}[\phi^-,\chi^-_{\rm
a}]\rangle_0 - \big[\langle S_{\rm int}[\phi^-,\chi^-_{\rm
a}]\rangle_0\big]^2\}.\nonumber
\end{eqnarray}
As a further step we make a separation of the $\phi$-field itself
into its long wavelength modes with $|k|<\mu$, which determine the
domains, and the short wavelength modes ($|k|>\mu$) which act as
their implicit environment. However, in calculating $t_D$ this
latter step only serves to further reduce decoherence times which,
as we shall see, are already short enough.

In reality, even if we ignore this step for the purpose of
bounding $t_D$ there is not a unique bound, since it depends on
the class of classical solutions considered. However, in practice
all sensible choices seem, qualitatively, to give the same upper
bound on $t_D$. The reasons for this are the following. Firstly,
for the long wavelength system modes the spatial profile of the
classical solution is not particularly relevant, since it is its
exponential growth in time that sets the scale for the onset of
classical behaviour. Thus, not only is $t_D$ insensitive to
wavelength for small-k modes, but it only depends logarithmically
on the parameters of the theory. Further, with the common
mass-scale $\mu$ that we assume here, the Compton wavelength
$\mu^{-1}$, both determines the rate of exponential growth and
provides the natural distance scale over which we do not wish to
discriminate between classical solutions. Different choices lead
to differences in $\mu t_D$ of order unity, significantly smaller
than $\mu t_D$ itself, which we ignore. For weak couplings it is
relatively easy to compute the upper bound on this time due to the
interactions $S_{\rm int}[\phi ,\chi ]$ of (\ref{Sint}). We shall
now see that, in our particular model, it is in general shorter
than the time $t^*$ at which the transition is complete, defined
in terms of the system field as the time for which
\begin{equation}
\langle \phi^2\rangle_{t^*}\sim \eta^2 = 6\mu^2/\lambda\,\,\,
,\label{deftsp}
\end{equation}
where the average is taken over the system field. That is, in
(\ref{deftsp}) we ignore the contributions from the dominant
environment given by the short wavelength ($|k|
>\mu$) modes of $\phi$. This renders $\langle \phi^2\rangle_{t^*}$ finite.
(More explicitly, in the language of our earlier papers based on
the master equation, e.g. \cite{npb}, $ \langle
\phi^2\rangle_{t^*} = Tr\{\rho_{\rm r}\phi^2\}, $ where $\rho_{\rm
r}$ is the {\it reduced} density matrix obtained on tracing out
the environment.) Any ambiguities are again of order unity in $\mu
t_D$, the timelag for non-linear effects to occur, in comparison
to the exponential runaway of the free field \cite{karra}.

 Since the effect of the
environment is to induce damping, the classical behaviour of the
field is expressed through the classical Langevin stochastic
equations that it satisfies \cite{nunoray,nunobett}. We are
assuming that coupling is sufficiently strong for the system not
to recohere after $t^*$ \cite{nuno}.

For the biquadratic coupling of Eq.(\ref{Sint}), the IF is given
by
\[
{\rm Re} \delta A = \frac{g^2}{8} \int d^4 x\int d^4y ~ \Delta (x)
K(x-y) \Sigma (y), \]
\[
{\rm Im} \delta A = - \frac{g^2}{16} \int d^4x\int d^4y ~ \Delta
(x) N (x,y) \Delta (y), \]
where $K(x-y) = {\rm Im} G_{++}^2(x,y) \theta (y^0-x^0)$ is the
dissipation kernel and $ N(x-y) = {\rm Re} G_{++}^2(x,y)$ is the
noise (diffusion) kernel. $G_{++}$ is the relevant
closed-time-path correlator of the $\chi$-field at temperature
$T_0$ \cite{npb}. We have defined $\Delta ={1\over{2}}(\phi^{+2} -
\phi^{-2})$ and $\Sigma ={1\over{2}} (\phi^{+2} + \phi^{-2})$.

We look for classical solutions of the form

\[
\phi_{\rm cl}(\vec x, s) =  f(s,t)\Phi(\vec x),\] where, in
principle, $f(s,t)$ satisfies $f(0,t)= \phi_{\rm i}$ and $f(t,t) =
\phi_{\rm f}$ and $\Phi(\vec x)$ gives the space-field
configuration.

We begin by showing that, using (\ref{tD2}), we recreate the
results obtained previously in Refs.\cite{lomplb,npb} on using the
more complicated master equation, in which the field is spread
through space, and decoherence is due to different field
amplitudes. In anticipation that $t_D<t^*$, it is sufficient to
restrict ourselves to the initial Gaussian free field evolution
with negative $\rm (mass)^2$.  In fact, for idealised Langevin
equations, this can be a good approximation for domain formation
into the non-linear regime \cite{moro}. We look for classical
classical fields that, after a sudden quench, have the form
\cite{lomplb}

\begin{equation}
\phi_{\rm cl}^{\pm}(s,\vec x)= e^{\mu s}\phi_{\rm f}^{\pm}
\cos(k_0\, x)\cos(k_0\, y)\cos(k_0\, z),\label{cboard}
\end{equation}
where $\phi_{\rm f}^\pm$ is the final field configuration. This is
a single mode approximation to a regular chequer-board domain
structure. Shorter wavelengths can be introduced without altering
the result significantly.  The reader is referred to \cite{npb}
for more details. For an instantaneous quench, we will use the
late time behaviour of the longest wavelengths ($k_0=0$),
$\phi_{\rm cl}^{\pm}(s,\vec x)\sim e^{\mu s}\phi_{\rm f}^{\pm}$.
The exponential factor, as always, arises from the growth of the
unstable long-wavelength modes.

Thus, ${\rm Im}\delta A[\phi^+_{\rm cl},\phi^-_{\rm cl}]$ takes
the form
\begin{eqnarray}
{\rm Im}~\delta A &=& \frac{g^2 V T_c^2\pi}{64} \Delta_{\rm f}^2
\int_0^\infty \frac{dk}{(k^2 + \mu^2)^2} \nonumber \\
&\times &\frac{1 + e^{4\mu t} - e^{2\mu t} \cos{(2\sqrt{k^2 +
\mu^2})}}{(k^2 + 2 \mu^2)}, \label{ImS}
\end{eqnarray}
where $\Delta_{\rm f} = 1/2 (\phi_{\rm f}^{+2} - \phi_{\rm
f}^{-2})$, and $T_c$ is the critical environmental-temperature. As
we noted in previous publications, the volume $V$ is due to the
fact we are considering field configurations spread over all
space. $V$ is interpreted as the minimal volume inside which there
are no coherent superpositions of macroscopically distinguishable
states for the field. Later, we shall consider localized
configurations where this factor does not appear.

After assuming $\mu t \gg 1$, the integral in momenta can be done
analytically obtaining,
\begin{equation}
{\rm Im}~\delta A \sim \frac{g^2 V T_c^2\pi^2}{256}\frac{(3 -
2\sqrt{2})}{\mu^3} \Delta_{\rm f}^2 e^{4\mu t}. \label{ImS2}
\end{equation}
With this expression at hand, we are able to evaluate the
decoherence time $t_D$ for amplitude variation as
\begin{equation}
\mu t_D \sim \frac{1}{2} \ln\left\{\frac{16 \mu^{\frac{3}{2}}}{g
T_c \Delta_{\rm f} V^{\frac{1}{2}}\pi\sqrt{(3 -
2\sqrt{2})}}\right\}.
\end{equation}

Using a conservative value for the volume, $V = {\cal
O}(\mu^{-3})$, we get
\begin{equation}
\mu t_D \sim \frac{1}{2} \ln\left\{\frac{16 \mu^{3}}{g T_c
\Delta_{\rm f}}\right\}\label{mutD}.
\end{equation}
We re-write last expression in terms of $\Delta_{\rm f} = \bar\phi
\bar\Delta /2$, with $\bar\phi = \phi_{\rm f}^+ + \phi_{\rm f}^-$
and $\bar\Delta = \phi_{\rm f}^+ - \phi_{\rm f}^-$. At the
completion of the transition ${\bar\phi}^2 \simeq \eta^2 \sim
\lambda^{-1}$, and we will adopt, at time $t_D$, ${\bar\phi}^2
\sim {\cal O}(\mu^2\alpha/\lambda )$. $\lambda < \alpha < 1$ is to
be determined self-consistently from the condition that, at time
$t_D$, $\langle\phi^2\rangle_t \sim \alpha\eta^2$. We have shown
in Ref.\cite{lomplb} that the value of $\alpha$ is determined as
$\alpha \approx \sqrt{\mu/T_c}$. We also set $\bar\Delta \sim 2
\mu$ (i.e. we do not discriminate between field amplitudes which
differ by ${\cal O}(\mu)$), where $\mu^{-1}$ characterizes the
thickness of domain boundaries (walls) as the field settles into
its ground-state values. Additionally, for simplicity we take the
couplings $g \sim \lambda$. Therefore, we obtain an upper bound on
$t_D$,

\begin{equation}
\mu t_D \sim \frac{1}{2}
\ln\left\{\frac{\eta}{T_c\sqrt{\alpha}}\right\},
\end{equation}
which exactly coincides with the result found in Ref.\cite{lomplb}
from the master equation for a sudden quench but, in the present
case, using the decoherence functional approach.

If we now trace over short-wavelength modes, as in
\cite{lombmazz}, but for unstable modes, we would get a further
term in ${\rm Im}~\delta A $ qualitatively similar to
(\ref{ImS2}), which will serve to preserve $t_D < t^*$ by making
$t_D$ even smaller. We shall not consider this implicit
environment in subsequent analysis.

For comparison, we find $t^*$, for which $\langle\phi^2\rangle_t
\sim \eta^2$, given by \cite{npb,lomplb}
\begin{equation}
\mu t^* \sim \frac{1}{2} \ln\left\{\frac{\eta}{\sqrt{\mu
T_c}}\right\},
\end{equation}
whereby $\mu^{-1} < t_D < t^*$, with

\begin{equation}
\mu t^* - \mu t_D \simeq \frac{1}{4}
\ln\left\{\frac{T_c}{\mu}\right\} > 1, \label{tD3}
\end{equation}
for weak enough coupling, or high enough initial temperatures.

Having seen that the decoherence functional approach gives
identical conclusions to the solution of master equations, the new
work in this letter is to evaluate the IF for different field
configurations which, from the point of view of the Master
equation, would be much more taxing analytically. Instead of the
classical solutions used before \cite{npb,proceed1}, in which the
field is spread in an infinite chequer-board through space, here
we are concerned with a different field configuration; a localized
domain wall.

The orientation of such a wall is irrelevant, as was the
orientation of the chequer-board solution used earlier. Most
simply,  we consider classical domain wall solutions (for the
$k_0=0$ mode) of the form
\begin{equation}
\phi^\pm (s,\vec x) = \phi_{\rm f}^\pm e^{\mu s}\tanh(\mu
x),\label{kink}
 \end{equation}
 which link adjacent domains.

Our main new result is to determine the decoherence induced by a
small displacement in the domain wall, by evaluating the influence
action for the classical field configurations $\phi^\pm (x,s) =
\eta f(s,t) \Phi^\pm (x)$, where $\Phi^\pm (x) = \Phi (x \pm
\delta/2)$. We will consider $\delta$ as a small displacement in
the position of the wall, and consequently we expand the classical
solution (or, equivalently, $\Delta (x)$) in powers of $\mu
\delta$, up to linear order

\[
\Delta(s,\vec x) \approx \mu ~\delta ~\eta^2 ~e^{2\mu s}
~\tanh(\mu x) ~{\rm sech}^2(\mu x).
\]

Doing this, the imaginary part of the influence action can be
written as (after integrating over time and assuming $\mu t \gg
1$),
\begin{eqnarray}
&&{\rm Im}\delta A \approx
\frac{g^2T_c^2\eta^4\delta^2\mu^2}{64(2\pi)^6}~ e^{4\mu t}~\int
d^3x
\int d^3y \nonumber \\
&\times &\int d^3k \int d^3p  ~~\frac{e^{-i(\vec p + \vec k)(\vec
x - \vec
y)}}{(k^2 + \mu^2)(p^2 + \mu^2)}  \nonumber \\
&\times &\frac{\tanh(\mu x)~{\rm sech}^2(\mu x)~\tanh(\mu y)~{\rm
sech}^2(\mu y)}{\left[(\sqrt{k^2 + \mu^2}+\sqrt{p^2 + \mu^2})^2 +
4 \mu^2\right]}\nonumber .\end{eqnarray}

These integrations can be exactly evaluated (in part analytically
and in part numerically), to give
\begin{equation}
{\rm Im}\delta A \approx 0.2 \frac{g^2T_c^2\eta^4
L^2\delta^2}{1024} \frac{e^{4\mu t}}{\mu^2}, \label{imkinks}
\end{equation}
where $L^2$ is a surface term, analogous to the volume $V$ in the
chequer-board analysis. This coefficient comes from the fact that
we are considering a one dimensional kink solution embedded in a
three dimensional space. The $L^2$ coefficient comes from the
``free surface'' or wall in two directions of the
three-dimensional kink. Considering very conservative values (such
as to obtain upper bounds for $t_D$), we can set $L ={\cal
O}(\mu^{-1})$ as the minimum length scale in which there could be
not coherent superpositions of macroscopic states of the field.

If the decoherence time due to displacements is ${\bar t}_D$, let
us suppose that $\mu {\bar t}_D
> 1$. Then ${\bar t}_D$ can be evaluated from the last equation, and its
order of magnitude is given by
\begin{equation}
\mu {\bar t}_D \sim \frac{1}{2} \ln\left\{\frac{74\mu}{g T_c
\eta^2 L \delta}\right\}.
\end{equation}
Superficially, this result looks similar to previous results,
where the difference between the field amplitudes has been
replaced by the scale of displacement suffered by the domain
boundary wall, $\delta$. The main difference is in the power of
$\eta$ inside the log. If one takes $\delta = \gamma \mu^{-1}$, we
may bound ${\bar t}_D$ as

\begin{equation}
\mu {\bar t}_D \sim \frac{1}{2} \ln\left\{\frac{12\mu}{\gamma
T_c}\right\}.
\end{equation}


Since $T_c \gg \mu$ (in fact one can show $T_c^2/\mu^2 \sim
24/\lambda$ \cite{npb}), this result puts a bound on the possible
value of $\gamma$ in order to have $\mu {\bar t}_D > 1$, i.e.
$\gamma \leq \mu/T_c\ll 1$, and hence $\delta\ll \mu^{-1}$. If
$\gamma$ is larger, at very early times in our model $t \sim
\mu^{-1}$, we get from Eq.(\ref{imkinks}) ${\rm Im}~\delta A
> 1$ immediately. Thus, the
system behaves classically from this earliest permissable time,
$\mu {\bar t}_D = {\cal O}(1)$.

Since the size of the core of the domain wall is ${\cal
O}(\mu^{-1})$, we take $\delta \sim {\cal O}(\mu^{-1})$ (i.e.
$\gamma = {\cal O}(1)$) as the minimum length scale in which there
could be not coherent superpositions of macroscopic states of the
field. With decoherence occurring at the earliest possible time it
follows automatically that
\begin{equation}
\mu t^* - \mu {\bar t}_D > 0.
\end{equation}
Even if we demand decoherence at scales a fraction of the domain
wall core thickness (i.e. $\gamma \ll 1$) we can ensure that
\begin{equation}
\mu t^* - \mu {\bar t}_D \sim \frac{1}{4}
\ln\left\{\frac{\gamma^2T_c}{\lambda \mu}\right\}  > 0.
\end{equation}
This result implies that decoherence due to the displacement of
the boundary is a very early time event ${\bar t}_D < t_D$.

The outcome of this analysis is that configurations of field
domains with displaced boundaries are less important for
decoherence than fixed domains in which there are field
fluctuations, which decohere later.

What is more appealing is, if one perform all the calculation in
1+1 dimensions, the coefficient $L$ appears only in our first
example (i.e. the plane-wave); not for the kink-like solutions.

In fact, rather than consider a whole chequer-board of domains as
in \cite{npb,lomplb,proceed1}, we can restrict ourselves to two
adjacent domains with boundary given by (\ref{kink}). Consider the
simple case where the difference between walls is given by
$\phi_{\rm f}^+ = \phi_{\rm f}^- - \epsilon$, with $\epsilon$ a
small fluctuation around the final value of the field
configuration. We assume that the final field configuration is
$\phi_{\rm f}^+ = \sqrt{\alpha}\eta$, with $\alpha$ the
self-consistent coefficient shown earlier and $\epsilon ={\cal
O}(\mu )$. Perhaps surprisingly, we recreate the result of
(\ref{tD3}) exactly.  Yet again, we have decoherent behaviour
before the transition is completed.

In addition, the field possesses classical correlations at early
times by virtue of the quasi-Gaussian nature of the regime
\cite{lomplb2,diana} to give a fully classical picture by time
$t^*$.

Finally, we return to the other view of domain boundaries
mentioned in the introduction, as topological defects. What does
this analysis have to say about the classical behaviour of defects
like vortices, whose separation also measures the size of domains
in simple circumstances \cite{kibble2}? Treating vortices as
having their cores as line zeroes (in the same way that the cores
of our domain walls are sheet zeroes) we have shown elsewhere that
the mechanism for the production of classical vortices has several
parts \cite{lomplb2}. Yet again, the environment renders the
long-wavelength modes of the order parameter field classical at
early times, by or before the transition is complete. In
particular, those on the scale of the separation of the
line-zeroes that will characterise the classical domains will have
decohered by the time the transition is complete, even though the
field modes on the scale of classical vortex thickness do not
decohere ever. For line-zeroes to mature into vortex cores the
field needs to have an energy profile commensurate with the vortex
solutions to the ordinary classical Euler-Lagrange equations. This
requires that the field fluctuations are peaked around
long-wavelengths, to avoid fluctuations causing wiggles in the
cores and creating small vortex loops, a related condition
satisfied in our models. The resultant density of line-zeroes can
already be inferred in the linear regime, whose topological
charges are well-defined even though close inspection of their
interior structure is not permitted classically. What the analysis
of this letter shows is that we expect decoherence due to vortex
displacement to be irrelevant in comparison to decoherence due to
field fluctuations.

The work of F.L and P.V was supported by UBA, CONICET, and ANPCyT,
Argentina. P.V specially thanks partial support from IUPAP.


\begin{thebibliography}{99}



\bibitem{npb}  F.C. Lombardo, F.D. Mazzitelli, and R.J. Rivers, Nucl. Phys.
{\bf B672}, 462 (2003)



\bibitem{lomplb}  F.C. Lombardo, F.D. Mazzitelli, and R.J. Rivers, Phys. Lett.
{\bf B523}, 317 (2001)



\bibitem{proceed1} R.J. Rivers and F.C. Lombardo, Int. J. Theor. Phys. {\bf 44}, 1855 (2005)
; R.J. Rivers and F.C. Lombardo, Brazilian Journal of Physics 35,
397 (2005)



\bibitem{guthpi} A. Guth and S.Y. Pi, Phys. Rev. {\bf D32}, 1899 (1991)



\bibitem{kibble1} T.W.B. Kibble, J. Phys. {\bf A9}, 1387
(1976)



\bibitem{zurek1} W.H. Zurek, Physics Reports {\bf 276}, 177, (1996)



\bibitem{lomplb2}  R.J. Rivers, F.C. Lombardo, and F.D. Mazzitelli, Phys.
Lett. {\bf B539}, 1 (2002)




\bibitem{POFer} F.C. Lombardo and D. L\'opez Nacir, Phys. Rev. {\bf D72}, 063506 (2005);
F.C. Lombardo, Brazilian Journal of Physics 35, 391 (2005)



\bibitem{lombmazz} F.C. Lombardo and F.D. Mazzitelli, Phys. Rev. {\bf D53}, 2001 (1996)

\bibitem{huzhang} B.L. Hu, in {\it Relativity and Gravitation: Classical and Quantum}, J.C.D.
Olivo et al (Eds.), World Scientific, Singapore (1991)






\bibitem{boya}  D. Boyanovsky, H.J. de Vega, and R. Holman, Phys.
Rev. {\bf D49}, 2769 (1994); D. Boyanovsky, H.J. de Vega, R.
Holman, D.-S. Lee, and A. Singh, Phys. Rev. {\bf D51}, 4419
(1995); S.A. Ramsey and B.L. Hu, Phys. Rev. {\bf D56}, 661 (1997)


\bibitem{moro} E. Moro and G. Lythe, Phys.Rev. {\bf E59} R1303 (1999).

\bibitem{Gri} R.B. Griffiths, J. Stat. Phys. {\bf 36}, 219 (1984)



\bibitem{Omn} R. Omnes, J. Stat. Phys. {\bf 53}, 893 (1988); Ann. Phys. {\bf 201}, 354 (1990); Rev. Mod.
Phys. {\bf 64}, 339 (1992)



\bibitem{GH} M. Gell-Mann and J.B. Hartle, Phys. Rev. {\bf D47}, 3345 (1993); J.J. Halliwell, Phys. Rev.
{\bf D60}, 105031 (1999)



\bibitem{calhu} E. Calzetta and B.L. Hu, Phys. Rev. {\bf D35}, 495 (1987)



\bibitem{feynver} R. Feynman and F. Vernon, Ann. Phys. (N. Y.) 24, 118 (1963)



\bibitem{calhumaz}E.A. Calzetta, B.L. Hu and F.D. Mazzitelli, Phys. Rep. {\bf 352}, 459 (2001)

\bibitem{karra} G. Karra and R.J. Rivers, Phys.Lett. B {\bf 414}, 28 (1997)

\bibitem{nunoray} N.D. Antunes, P. Gandra, R.J. Rivers, and A. Swarup, Phys. Rev. {\bf D73}, 085012 (2006);
 N.D. Antunes, P. Gandra, and R.J. Rivers, Phys. Rev. {\bf D71}, 105006 (2005)



\bibitem{nunobett} N.D. Antunes, L.M.A. Bettencourt, and W.H. Zurek, Phys. Rev. Lett. {\bf 82}, 2824 (1999)


\bibitem{nuno} N.D. Antunes, F.C. Lombardo and D.  Monteoliva, Phys. Rev.
{\bf E64}, 066118 (2001);  N.D. Antunes, F.C. Lombardo, D.
Monteoliva, and P.I. Villar, Phys. Rev. {\bf E73}, 066105 (2006)




\bibitem{diana} F.C. Lombardo, F.D. Mazzitelli, and D. Monteoliva, Phys. Rev.
{\bf D62}, 045016 (2000)


\bibitem{kibble2} T.W.B. Kibble, Phys. Rep. {\bf 67}, 183 (1980)









\end{thebibliography}
\end{document}